\documentstyle[psfig]{mn}
\newif\ifAMStwofonts
\ifoldfss
  \ifCUPmtlplainloaded \else
    \NewTextAlphabet{textbfit} {cmbxti10} {}
    \NewTextAlphabet{textbfss} {cmssbx10} {}
    \NewMathAlphabet{mathbfit} {cmbxti10} {} 
    \NewMathAlphabet{mathbfss} {cmssbx10} {} 
  \fi
  \ifAMStwofonts
    \ifCUPmtlplainloaded \else
      \NewSymbolFont{upmath} {eurm10}
      \NewSymbolFont{AMSa} {msam10}
      \NewMathSymbol{\upi}     {0}{upmath}{19}
      \NewMathSymbol{\umu}     {0}{upmath}{16}
      \NewMathSymbol{\upartial}{0}{upmath}{40}
      \NewMathSymbol{\leqslant}{3}{AMSa}{36}
      \NewMathSymbol{\geqslant}{3}{AMSa}{3E}

       \let\le=\leqslant
       \let\ge=\geqslant
    \fi
  \fi
\fi 

\ifnfssone
  \newmathalphabet{\mathit}
  \addtoversion{normal}{\mathit}{cmr}{m}{it}
  \addtoversion{bold}{\mathit}{cmr}{bx}{it}
  \newmathalphabet{\mathbfit} 
  \addtoversion{normal}{\mathbfit}{cmr}{bx}{it}
  \addtoversion{bold}{\mathbfit}{cmr}{bx}{it}
  \newmathalphabet{\mathbfss} 
  \addtoversion{normal}{\mathbfss}{cmss}{bx}{n}
  \addtoversion{bold}{\mathbfss}{cmss}{bx}{n}
  \ifAMStwofonts
    \ifCUPmtlplainloaded \else
      \UseAMStwoboldmath
      \makeatletter
      \new@mathgroup\upmath@group
      \define@mathgroup\mv@normal\upmath@group{eur}{m}{n}
      \define@mathgroup\mv@bold\upmath@group{eur}{b}{n}
      \edef\UPM{\hexnumber\upmath@group}
      \new@mathgroup\amsa@group
      \define@mathgroup\mv@normal\amsa@group{msa}{m}{n}
      \define@mathgroup\mv@bold\amsa@group{msa}{m}{n}
      \edef\AMSa{\hexnumber\amsa@group}
      \makeatother
      \mathchardef\upi="0\UPM19
      \mathchardef\umu="0\UPM16
      \mathchardef\upartial="0\UPM40
      \mathchardef\leqslant="3\AMSa36
      \mathchardef\geqslant="3\AMSa3E

       \let\le=\leqslant
       \let\ge=\geqslant
    \fi
  \fi
\fi 

\ifnfsstwo
  \DeclareMathAlphabet{\mathbfit}{OT1}{cmr}{bx}{it}
  \SetMathAlphabet\mathbfit{bold}{OT1}{cmr}{bx}{it}
  \DeclareMathAlphabet{\mathbfss}{OT1}{cmss}{bx}{n}
  \SetMathAlphabet\mathbfss{bold}{OT1}{cmss}{bx}{n}
  \ifAMStwofonts
    \ifCUPmtlplainloaded \else
      \DeclareSymbolFont{UPM}{U}{eur}{m}{n}
      \SetSymbolFont{UPM}{bold}{U}{eur}{b}{n}
      \DeclareSymbolFont{AMSa}{U}{msa}{m}{n}
      \DeclareMathSymbol{\upi}{0}{UPM}{"19}
      \DeclareMathSymbol{\umu}{0}{UPM}{"16}
      \DeclareMathSymbol{\upartial}{0}{UPM}{"40}
      \DeclareMathSymbol{\leqslant}{3}{AMSa}{"36}
      \DeclareMathSymbol{\geqslant}{3}{AMSa}{"3E}

       \let\le=\leqslant
       \let\ge=\geqslant
    \fi
  \fi
\fi 

\ifCUPmtlplainloaded \else
  \ifAMStwofonts \else 
    \def\upi{\pi}
    \def\umu{\mu}
    \def\upartial{\partial}
  \fi
\fi

\title[Hall drift in neutron stars]{The influence of Hall drift
 on the magnetic fields of neutron stars}
\author[R. Hollerbach and G. R\"udiger]
       {Rainer Hollerbach$^1$\thanks{Permanent address: Department of
  Mathematics, University of Glasgow, Glasgow G12 8QW, United Kingdom,
  rh@maths.gla.ac.uk}
        and G\"unther R\"udiger$^2$\\
        $^1$Brandenburgische Technische Universit\"at Cottbus,
            LAS, Fakult\"at 3, 03013 Cottbus, Germany\\
        $^2$Astrophysikalisches Institut Potsdam,
            An der Sternwarte 16, 14482 Potsdam, Germany}

\date{Accepted 2002 July 24.
      Received 2002 June 17;
      in original form 2002 March 15}

\pagerange{\pageref{firstpage}--\pageref{lastpage}}
\pubyear{2002}

\begin{document}

\maketitle

\label{firstpage}

\begin{abstract}
We consider the evolution of magnetic fields under the influence of Hall drift
and Ohmic decay.  The governing equation is solved numerically, in a spherical
shell with $r_i/r_o=0.75$.  Starting with simple free decay modes as initial
conditions, we then consider the subsequent evolution.  The Hall effect
induces so-called helicoidal oscillations, in which energy is redistributed
among the different modes.  We find that the amplitude of these oscillations
can be quite substantial, with some of the higher harmonics becoming comparable
to the original field.  Nevertheless, this transfer of energy to the higher
harmonics is not sufficient to significantly accelerate the decay of the
original field, at least not at the $R_B=O(100)$ parameter values accessible
to us, where this Hall parameter $R_B$ measures the ratio of the Ohmic
timescale to the Hall timescale.  We do find clear evidence though of
increasingly fine structures developing for increasingly large $R_B$,
suggesting that perhaps this Hall-induced cascade to ever shorter lengthscales
is eventually sufficiently vigorous to enhance the decay of the original field.
Finally, the implications for the evolution of neutron star magnetic fields
are discussed.
\end{abstract}


\section{Introduction}

Neutron stars have the strongest magnetic fields found in the
universe, with fields perhaps as large as $10^{15}$ G for so-called magnetars
(e.g.\ Murakami 1999), around $10^{12}$ G for young ($\sim10^7$ year) radio
and X-ray pulsars, and a still appreciable $10^8 - 10^{10}$ G for much older
($\sim10^{10}$ year) millisecond pulsars (e.g.\ Chanmugam 1992; Bhattacharya
1995; Lyne 2000).  This correlation between field strength and age suggests
that these very different strengths are due to the field decaying in time,
rather than to any intrinsic differences between different neutron stars.
One would therefore like to identify the processes causing the field to decay.

The additional observation that most weakly magnetic neutron stars have binary
companions, whereas very few strongly magnetic ones do (e.g.\ Bhattacharya
1995), suggests that accretion of mass from the companion is somehow causing
the field to decay (by mechanisms that need not concern us here, but see for
example Blondin \& Freese 1986; Romani 1990; Urpin \& Geppert 1995).  The
observational evidence is unfortunately inconclusive, with Taam \& van den
Heuvel (1986) claiming a correlation between field strength and accreted mass,
but Wijers (1997) disputing this.

One would therefore like to consider the possibility of other
mechanisms besides accretion.  One such alternative is Hall drift, first
proposed by Jones (1988), in which the magnetic field influences itself
through a quadratic nonlinearity.  If it is relevant at all, Hall drift will
therefore be most important for the very strongest fields -- which as we saw
tend to occur in isolated neutron stars, where accretion is not acting at all.
Hall drift is thus likely to be the dominant mechanism influencing the
magnetic fields of these stars.  Of course, it could potentially be important
in binaries as well, at least in the early stages while their fields are still
relatively strong.

Again as a result of this quadratic nonlinearity, the timescale on which Hall
drift might be expected to act is almost necessarily inversely proportional to
$|{\bf B}|$.  Jones suggests that it is given by
\begin{equation}
t_{Hall}\sim {10^8\over B_{12}}\;{\rm years},
\end{equation}
where $B_{12}$ is the field strength in units of $10^{12}$ G.  See also
Goldreich \& Reisenegger (1992), who obtain a similar estimate.
For these $O(10^{12})$ G radio pulsars, one therefore expects a
timescale comparable to their age.
And indeed, Lyne, Manchester \& Taylor (1985) and Narayan \& Ostriker (1990)
have suggested that the fields of these young pulsars do decay on a $10^7$
year timescale (although this too is in dispute, see for example Hartman et
al.\ 1997; Regimbau \& Pacheco 2001).  Apart from the strength of the
observational evidence though, the mere fact that Hall drift could affect the
fields of neutron stars on timescales so short compared to their evolutionary
timescales makes it worthy of study.

There is one slight difficulty though in attributing this possible
$10^7$ year decay rate to Hall drift, namely that the Hall effect conserves
magnetic energy, and therefore by itself cannot cause any field decay at all.
The suggestion therefore is that the Hall term, being nonlinear, will
redistribute energy among the different modes, and in particular will initiate
a cascade to ever shorter lengthscales, where ordinary Ohmic decay (which only
acts on $O(10^{10})$ year timescales at the longer lengthscales) can destroy
the field.

Since this mechanism was first proposed by Goldreich \& Reisenegger, detailed
calculations have been done by a number of authors, including Naito \& Kojima
(1994), Muslimov (1994), Muslimov, Van Horn \& Wood (1995), Shalybkov \&
Urpin (1997) and Urpin \& Shalybkov (1999).  Of these, only the last two were
in the astrophysically relevant limit of large Hall parameter $R_B$ though,
where $R_B$ measures the ratio of the Ohmic timescale to the Hall timescale,
and is defined more precisely below.  However, it is not certain whether their
results were fully resolved, as they had only 20 radial by 40 latitudinal
finite difference points.

In contrast, we have 25 radial by 100 latitudinal
spectral expansion functions, and obtain fully resolved solutions for $R_B$
up to $O(100)$ (comparable to what Shalybkov \& Urpin achieved, and indeed in
broad agreement with their results, suggesting that perhaps their resolution
was good enough after all to resolve the most important features anyway).
Even at these large values, however, we find that while the Hall effect does
indeed induce a significant redistribution of energy among the different
modes, it does not appear to be enough to cause the lowest modes to decay
substantially faster than they would have otherwise.

Before applying this conclusion to real neutron stars though, it is important
to qualify it by noting that our calculations (as well as the others cited
above) are restricted to $\bf B$ being axisymmetric, and the various material
properties such as the density being independent of depth.  Neither of these
assumptions holds in real neutron stars, and relaxing either could
significantly alter the results.  For example, Vainshtein, Chitre \& Olinto
(2000) show that including variations in density introduces new effects even
for field configurations where no ordinary Hall drift would be present at all.
Similarly, Rheinhardt \& Geppert (2002) also consider field configurations
where no ordinary Hall cascade is present, but claim that instabilities,
including non-axisymmetric ones, can nevertheless arise.  We will discuss both
of these papers more fully below, as well as how these two restrictions might
be relaxed in future work.

\section{Equations, etc.}

\subsection{The Evolution Equation for $\bf B$}

The equation governing the evolution of a magnetic field under the influence
of Hall drift and Ohmic decay is
\[
{\partial{\bf B}\over\partial t}=-\nabla\times\Bigl({c\over4\pi ne}\,
    {(\bf\nabla\times B)\times B}\Bigr)
\]
\begin{equation}
\phantom{{\partial{\bf B}\over\partial t}=}
   - \nabla\times\Bigl({c^2\over4\pi\sigma}\,\nabla\times{\bf B}\Bigr),
\end{equation}
where $n$ is the electron number density, $\sigma$ the conductivity, $e$ the
electron charge and $c$ the speed of light.  See for example Goldreich \&
Reisenegger (1992) for the details of the derivation.  We then
nondimensionalize according to
\begin{equation}
{\bf B}=B_0\,{\bf B^*},\qquad l=r_o\,l^*,\qquad t=\tau\,t^*,
\end{equation}
where $B_0$ is a typical field strength, $r_o$ the radius of the star, and
\begin{equation}
\tau={4\pi ne r_o^2\over cB_0}
\end{equation}
the Hall timescale, where we note, incidentally, that these estimates (1)
amount to nothing more than inserting particular numbers into this result (4)
obtained purely by dimensional analysis.  We also note, though, that while
we've implicitly assumed $n$ to be constant here, in real neutron stars it
varies by several orders of magnitude throughout the depth of the crust.
It is therefore somewhat misleading to talk about a single Hall timescale;
there is rather a range of timescales, from perhaps $10^5/B_{12}$ up to
$10^8/B_{12}$ years.

Dropping the asterisks again, we obtain
\begin{equation}
{\partial{\bf B}\over\partial t}=-\nabla\times\bigl(
    {(\bf\nabla\times B)\times B}\bigr)
   + R_B^{-1}\,\nabla^2{\bf B},
\end{equation}
where we've assumed $\sigma$ to be constant as well (again not the case in
real neutron stars).  The Hall parameter
\begin{equation}
R_B={\sigma B_0\over nec},
\end{equation}
and is up to $10^2$ to $10^3$ in the crusts of $O(10^{12})$ G neutron
stars, where we will apply (5).  One useful physical interpretation to
associate with $R_B$ is the ratio of the Ohmic timescale $4\pi\sigma
r_o^2/c^2$ to this Hall timescale $\tau$.

\subsection{Cascades?}

We note that this Hall equation (5) bears certain similarities to the
vorticity equation of ordinary fluid dynamics,
\begin{equation}
{\partial{\bf\Omega}\over\partial t}=\nabla\times\bigl(
    {\bf U\times \Omega}\bigr)
   + Re^{-1}\,\nabla^2{\bf\Omega},
\end{equation}
where $\bf U$ and $\bf\Omega=\nabla\times U$ are the velocity and vorticity,
respectively, and $Re$ the Reynolds number.  It is on the basis of these
similarities that Goldreich \& Reisenegger suggested that the Hall effect
would initiate a cascade to ever shorter wavelengths, analogous to the
Kolmogorov cascade in ordinary fluid turbulence.  By applying arguments from
turbulence theory, they went on to suggest that the power spectrum of Hall
turbulence should fall off like $k^{-2}$, where $k$ is the wavenumber, and
that the dissipation scale should be reached when $k=O(R_B)$.

However, there is also one crucial difference between (5) and (7), one that
we believe has perhaps not been sufficiently appreciated before.  In
particular, in (7) the nonlinear term contains only first derivatives of
$\bf\Omega$, whereas in (5) the nonlinear term contains second derivatives
of $\bf B$. This has at least two important consequences.

First, in (5) the coefficients of the second derivative terms then depend on
the solution itself, whereas in (7) they don't.  This raises the possibility
that the mathematical character of the Hall equation could switch, from
parabolic to hyperbolic.  What would happen then is not clear, but the effect
could be dramatic, given how different these two types of equations are.
See, for example, Ockendon et al.\ (1999) for the theory behind the
classification of partial differential equations into parabolic, hyperbolic,
or elliptic types, depending on the sign of certain combinations of the
coefficients of the second derivative terms.

Second, in (7) one can always be certain that if one just goes to sufficiently
short lengthscales, the diffusive term will eventually dominate the advective
term, regardless of how large $Re$ is.  In contrast, in (5) one can go to
arbitrarily short lengthscales, and still not be certain that the diffusive
term will dominate the Hall term, because they both scale quadratically with
the wavenumber.  That means though that the whole notion of a definite
dissipation scale is much less clear in (5) than in (7).  One obtains a
definite dissipation scale only if one simply assumes that the coupling is
purely local in wavenumber space.  The argument is essentially as follows:  

By definition, the dissipation scale
occurs when the local Hall parameter is $O(1)$.  What is the `local' Hall
parameter though, when the defining equation (6) doesn't involve lengthscales
at all??  Well, {\it if} the coupling is purely local in wavenumber space,
then implicitly (6) does involve lengthscales after all, since then the $B_0$
that should be used is the field at that wavenumber only, rather than the
total field.  That is, one has
\begin{equation}
R_B'=R_B \cdot (B'/B),
\end{equation}
where the primed quantities denote these small-scale, local values,
and the unprimed the large-scale, global.  If in addition one has a $k^{-2}$
power spectrum, then $B'/B\sim k^{-1}$, so $R_B'$ is indeed reduced to $O(1)$
when $k=O(R_B)$.  We can see though how crucially the argument depends
on the coupling being purely local in wavenumber space; if this does not hold
then $R_B'=R_B$, and one simply does not obtain a definite dissipation scale
at all.  (We note also that there is no reason why the spectrum should not just
drop off like $k^{-p}$ indefinitely; provided $p>1$ the total energy would
certainly still be bounded.)

\subsection{Instabilities?}

Another intriguing idea, intended precisely to explore this issue of whether
the coupling is purely local in wavenumber space, is due to Rheinhardt \&
Geppert (2002), who considered fields satisfying
\begin{equation}
\nabla\times\bigl({(\bf\nabla\times B)\times B}\bigr)=0,
\end{equation}
so that no ordinary Hall cascade is present.  Linearizing (5) about such a
basic state, one obtains
\[
{\partial{\bf b}\over\partial t}=-\nabla\times\bigl(
    {(\nabla\times {\bf b})\times {\bf B}_0}
  + {(\nabla\times {\bf B}_0)\times {\bf b}}\bigr)
\]
\begin{equation}
   \phantom{{\partial{\bf b}\over\partial t}=}
   + R_B^{-1}\,\nabla^2{\bf b}.
\end{equation}
Neglecting the Ohmic decay of the basic state ${\bf B}_0$, one therefore has a
simple eigenvalue problem for the growth or decay rates of the perturbations
$\bf b$.  Rheinhardt \& Geppert then found that for sufficiently large $R_B$
arbitrarily short lengthscales could still be excited, which, they claimed,
proves that the coupling is not purely local in wavenumber space.

The difficulty we have with this approach is that while these small scale
modes may indeed be excited, we do not believe they can be distinguished from
the action of the ordinary Hall cascade.  In particular, while these small
scale modes do grow, the fastest growing modes are always large scale.  As soon
as these are excited though, the field no longer satisfies (9), and will
therefore generate an ordinary cascade, which is likely to reach these small
scales well before these postulated instabilities become significant.  For
example, our integration times here are sufficiently short that these
instabilities should not be manifesting themselves, and yet we do obtain very
short lengthscales, suggesting that it is the cascade rather than the
instabilities that is most significant.  Indeed, once the cascade is
established, it makes little sense at all to consider the growth or decay
rates of isolated modes.  The problem is intrinsically nonlinear, and must
be solved as such.

Finally, one might just note that there is one type of instability that
could be unambiguously distinguished from the ordinary cascade, namely a
non-axisymmetric one.  Here we will consider only axisymmetric solutions, so
the issue does not arise, but in general one might ask how one could go from
two-dimensional to three-dimensional solutions.  In particular, if one starts
with a 2D field, the ordinary cascade will forever remain 2D as well.  The only
way to obtain a 3D field is via an instability to a non-axisymmetric mode.  Of
course, as soon as one does have a 3D field, the cascade will also be 3D, so
one would again find it difficult to distinguish between the cascade and the
instability.  The initial trigger though that allows the field, and hence also
the cascade, to go from 2D to 3D would clearly have to be a non-axisymmetric
instability of some sort.

\subsection{T-P Decomposition}

Returning to our development of (5), for these axisymmetric fields we will
consider here, it is convenient to decompose $\bf B$ into toroidal and poloidal
components
\begin{equation}
{\bf B}={\bf B}_t + {\bf B}_p = B{\bf\hat e}_\phi
                   + \nabla\times(A{\bf\hat e}_\phi).
\end{equation}
Equation (5) then yields
\begin{equation}
{\partial A\over\partial t}=-{\bf\hat e}_\phi\cdot\bigl(
    (\nabla\times{\bf B}_t)\times{\bf B}_p\bigr)
    + R_B^{-1}D^2 A,
\end{equation}
\[
{\partial B\over\partial t}=-{\bf\hat e}_\phi\cdot\nabla\times\bigl(
    (\nabla\times{\bf B}_p)\times{\bf B}_p
   +(\nabla\times{\bf B}_t)\times{\bf B}_t\bigr)
\]
\begin{equation}
\phantom{BB=\,}   + R_B^{-1}D^2 B,
\end{equation}
where
\begin{equation}
D^2 f={1\over r}{\partial^2\over\partial r^2}\Bigl(fr\Bigr) + {1\over r^2}
  {\partial\over\partial\theta}\Bigl({1\over\sin\theta}
  {\partial\over\partial\theta}\bigl(f\sin\theta\bigr)\Bigr).
\end{equation}
We note then that if the field is initially purely toroidal it will remain
so, whereas if it is initially purely poloidal it will immediately induce a
toroidal part as well, and once both components are present each will act
back on the other.

\subsection{Boundary Conditions}

Taking the region exterior to the star to be a source-free vacuum, the
outer boundary conditions are simply
\begin{equation}
B=0,\qquad \Bigl({d\over dr}+{l+1\over r}\Bigr)\,A=0\qquad{\rm at\ }r=r_o,
\end{equation}
where $l$ is the spherical harmonic degree.  Since the numerical solution
already involves decomposition of $A$ and $B$ into Legendre functions,
implementing this $l$-dependent boundary condition presents no difficulties.

The inner boundary conditions are not quite so straightforward, and depend
very much on what assumptions we make about the interior of the star, about
which little is known for certain.  However, one common assumption
(e.g.\ Bhattacharya \& Datta 1996; Konenkov \& Geppert 2001) is that it is
superconducting, in which case the magnetic field will be expelled from it.
The boundary conditions are then that the normal component of the magnetic
field and the tangential components of the associated electric field must
vanish.  $B_r=0$ immediately yields $A=0$, but $E_t=0$ is a little more
complicated, and requires a little algebra before yielding
\begin{equation}
{1\over r\sin\theta}{\partial\over\partial\theta}\Bigl(B\sin\theta\Bigr)B
  + R_B^{-1}\,{1\over r}{\partial\over\partial r}\Bigl(Br\Bigr)=0.
\end{equation}
Such a nonlinear boundary condition is unfortunately very difficult to
implement.  We would therefore like to simplify it in some way.  We do so by
noting that in the relevant $R_B\gg1$ limit the second term ought to be
negligible (assuming $\partial B/\partial r$ does not increase with $R_B$,
that is, assuming that no boundary layers develop), in which case we are left
with just $B=0$.  For our inner boundary conditions we therefore take
\begin{equation}
B=0,\qquad A=0\qquad{\rm at\ }r=r_i.
\end{equation}

The radii at which we will apply these boundary conditions are $r_i=0.75$ and
$r_o=1$.  Although this is still not quite as thin as neutron star crusts are
believed to be ($r_i/r_o \approx 0.9$ would be more appropriate), it should
be enough to capture most of the geometrical effects of having a thin shell,
but without experiencing numerical difficulties due to too extreme a
disparity between the radial and latitudinal lengthscales.

Finally, a few runs were also done with insulating boundaries inside as well
as outside.  This is obviously not realistic, but allows one to assess the
extent to which the solutions are affected by differing boundary conditions.
It turned out that while this certainly altered the quantitative details, the
general features remained the same.

\subsection{Initial Conditions}

In many problems the specific initial conditions are largely irrelevant, as
one is only interested in the final, equilibrated solutions.  In this case
though, the only `equilibrated' solution is ${\bf B}=0$, since (as noted
above, and as we will show below), all solutions of (5) necessarily
decay in time.  So, what we are interested in instead is to start with some
particular initial condition, and study the precise manner of the decay,
whether it is significantly faster than just Ohmic decay, whether higher
harmonics are excited in the process, etc.  In this problem therefore we
need to give careful consideration to our choice of initial conditions.

If we temporarily neglect the Hall terms in equations (12) and (13), we can
solve for the individual free decay modes.  Figure 1 shows the lowest $l=1$
and $l=2$ poloidal modes, and also the lowest $l=2$ toroidal mode.  The free
decay rates for these three modes are $49R_B^{-1}$, $61R_B^{-1}$, and
$166R_B^{-1}$.  We label them ${\bf B}_{p1}$, ${\bf B}_{p2}$, and
${\bf B}_{t2}$, and normalize them so that $B_r(r_o,0)=1$ for the poloidal
modes, and $B_{\rm max}=1$ for the toroidal mode.  Our initial conditions
will then consist of either these modes in isolation, or else simple linear 
combinations of them.

\begin{figure}
\psfig{file=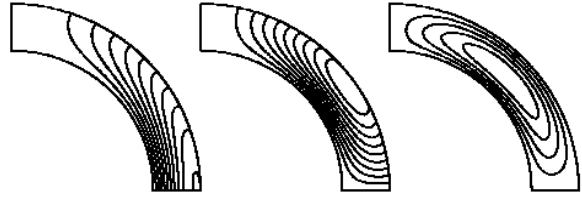,width=8.0cm}
\caption{Field lines of the two poloidal modes ${\bf B}_{p1}$ and
${\bf B}_{p2}$, and contours of the toroidal mode ${\bf B}_{t2}$.
${\bf B}_{p1}$ is equatorially symmetric, ${\bf B}_{p2}$ and ${\bf B}_{t2}$
antisymmetric.}
\end{figure}

\subsection{Symmetries}

At this point it is worthwhile also to briefly consider some of the symmetries
associated with (12) and (13), to avoid doing effectively duplicate runs.  For
example, (5) is clearly not invariant under ${\bf B}\rightarrow-{\bf B}$, so do
we need to consider $\pm{\bf B}_{p1}$, $\pm{\bf B}_{p2}$, and $\pm{\bf B}_{t2}$
separately?  Well, $\pm{\bf B}_{p1}$ are physically the same, just turned
upside down, so clearly not in that case.  For $\pm{\bf B}_{p2}$ though it's
not so obvious, since these are not the same; $+{\bf B}_{p2}$ has the field
inward in a ring around the equator, and outward at the poles, whereas
$-{\bf B}_{p2}$ has the reverse, and no amount of tilting or turning will
cause the two to coincide.  Nevertheless, one can see easily enough that
$\pm{\bf B}_{p2}$ will still evolve in exactly the same way, by noting that
(12) is linear in $A$, whereas (13) contains only even powers of
$A$.  Reversing the sign of {\it any} initial poloidal field will
therefore always yield exactly the same evolution.  Then, having noted how
$A$ enters into (12) and (13), and how that affects this $\pm$ symmetry, it is
an easy matter to verify that because $B$ enters differently, it does
not share this symmetry.  In general, therefore, reversing the sign of an
initial toroidal field will affect the evolution.  We will therefore have to
consider $\pm{\bf B}_{t2}$ separately.

Another symmetry worth mentioning is the equatorial symmetry, particularly as
this is somewhat different from that usually encountered in stellar dynamos.
One may readily verify that solutions exist having $A$ either symmetric
or antisymmetric, but with $B$ antisymmetric in both cases.  In contrast, in
stellar dynamo models the parity of $B$ would also change, always being
the opposite of $A$'s (e.g.\ Knobloch, Tobias \& Weiss 1998).  We can see
easily enough though that that cannot be the case here, by noting this same
property of (12) and (13) already used above, that $A$ enters linearly in (12),
and as even powers only in (13).  Therefore, if either pure parity is allowed
for $A$ at all, the opposite one must also be allowed, but with $B$ having the
same parity in both cases.  Being aware of these equatorial symmetries is
obviously helpful as well in doing the runs, as one can then reduce the
computational effort by a factor of two for many of them.

Finally, combining the plus/minus and equatorial symmetries, we note in
passing that if one has an initial poloidal field that is equatorially
asymmetric, one may reverse the sign of either the symmetric or the
antisymmetric parts separately, and will still obtain exactly the same
evolution, with the only effect on the toroidal field being to reverse
the sign of its symmetric part (which is consistent, of course, with this
part being absent if $A$ has a pure parity, and also consistent with $B$
being unchanged if both parts of $A$ are reversed).

\subsection{Numerical Solution}

The code we use to solve (12), (13), (15) and (17) is a suitably modified
version of the spherical harmonic code described by Hollerbach (2000), where
we include modes up to $l=100$, and 25 Chebychev polynomials in the radial
direction.  A few runs were also redone at truncations as low as $70\times20$,
and as high as $120\times30$, to check whether $100\times25$ is adequate.  We
believe that the solutions presented here are indeed fully resolved, although
one of them does show some unusual features in its power spectrum, as we will
discuss below.

We note also that at $100\times25$, we are capable of resolving structures as
fine as $r_o\Delta\theta=\pi/100=0.03$ in latitude, and $\Delta r=0.25/25=0.01$
in radius (strictly speaking probably only structures two or three times
greater in each case, to allow for the fact that two or three collocation
points are needed to resolve a given `structure').  Even though the truncation
in $r$ is lower, the resolution is therefore already higher.  The reason for
this is that in such a thin shell one might also expect finer structures to
develop in $r$, as will indeed turn out to be the case (although typically not
three times finer).

Finally, the timestep used was $\Delta t=10^{-7}$ for most runs, with again a
few done at even smaller values.  Such small values were necessary to avoid
numerical instabilities.  The origin of these instabilities is almost certainly
the previously noted feature that the Hall term (which is treated explicitly)
contains just as many derivatives as the Ohmic term (treated implicitly).  It
is certainly well known that treating second derivative terms explicitly almost
invariably requires extremely small timesteps, with the maximum allowable
$\Delta t$ also decreasing very quickly with increasing truncation, as was
found to be the case here.

Once again, also, this feature that the Hall term has just as many derivatives
as the Ohmic term raises the possibility that the governing equation (5) could
switch from parabolic to hyperbolic.  What would happen then is not clear, but
the code certainly could not cope with that.  And indeed, we will find that
there are limits beyond which we simply cannot push $R_B$, no matter how much
we reduce $\Delta t$, indicating that perhaps such a switch has occurred.

\subsection{The Magnetic Energy Equation}

As noted above, one major purpose of this work is to address the question
as to whether the Hall effect can significantly accelerate the decay of a
magnetic field despite the fact that by itself it conserves magnetic energy.
It therefore seems appropriate to verify that it really does so, and in
the process derive a useful diagnostic equation to help assess the accuracy
of our numerical solutions.  To obtain this equation for the magnetic energy,
we begin by taking the dot product of (5) with $\bf B$ and applying various
vector identities to obtain
\[
{\partial\over\partial t}{{\bf B}^2\over2} = -\nabla\cdot\bigl[{\bf
  (J\times B)\times B} + R_B^{-1}\,({\bf J\times B})\bigr]
\]
\begin{equation}
  \phantom{{\partial\over\partial t}{{\bf B}^2\over2} =}
  - R_B^{-1}\,{\bf J}^2,
\end{equation}
where $\bf J=\nabla\times B$.  When integrated over the shell, therefore,
the Hall term will contribute only surface integrals at $r_i$ and $r_o$,
whereas the diffusive term will contribute both surface integrals and a
negative-definite volume integral.

In order to obtain our desired result, we therefore need to consider these
various surface terms very carefully, particularly the ones at $r_i$, where
we remember our $B=0$ boundary condition is only a computationally convenient
approximation to the true condition (16).  If this approximation should turn
out to yield some spurious source or sink of energy through the boundary, we
would not be able to use it after all.  We must therefore show that
\begin{equation}
{\bf\hat e}_r\cdot\bigl[{\bf (J\times B)\times B}
  + R_B^{-1}\,({\bf J\times B})\bigr]=0\qquad{\rm at\ }r=r_i.
\end{equation}
To do so, we begin by noting that in terms of the individual components
$(B_r,B_\theta,B_\phi)$ and $(J_r,J_\theta,J_\phi)$,
\begin{equation}
A=B=0\quad\Longrightarrow\quad B_r=B_\phi=J_r=0,
\end{equation}
so that, using also the generally valid result $J_\phi=-D^2 A$,
\begin{equation}
{\bf J\times B}=(B_\theta\, D^2 A,0,0)\qquad{\rm at\ }r=r_i.
\end{equation}
Next, (12) can be expressed as
\begin{equation}
{\partial A\over\partial t}=J_\theta B_r - J_r B_\theta
    + R_B^{-1}D^2 A,
\end{equation}
so applied at $r_i$, where we remember $A=B_r=J_r=0$, we find that $D^2 A=0$
as well.  We therefore have that
\begin{equation}
{\bf J\times B}=0\qquad{\rm at\ }r=r_i,
\end{equation}
which establishes our required result (19).

In contrast, at $r_o$ one finds that these surface terms do not vanish.
Instead, they turn out to be precisely what is needed to take into account
changes in the energy stored in the external field.  The final result is
then
\begin{equation}
{\partial\over\partial t}\,{1\over2}\int{\bf B}^2\,dV =
  - R_B^{-1}\int{\bf J}^2\,dV,
\end{equation}
where the integral on the left extends over $r\ge r_i$, and the one on the
right over $r_i\le r\le r_o$.  Equation (24) is thus the desired energy
balance, namely that the total magnetic energy in all of space decreases only
as a result of Ohmic decay.  Hall drift rearranges the field, and
hence also the energy, but neither creates nor destroys it.

In addition to its role in illuminating the physics of Hall drift and Ohmic
decay, (24) is also a very useful diagnostic tool in assessing the accuracy
of our solutions.  Reassuringly, we found that not only does the magnetic
energy indeed decrease monotonically in time (hardly a very stringent test),
but that all of our runs satisfied (24) to within 1 per cent or better.
That is, if we ({\it a posteriori}) compute the quantity
\begin{equation}
q={\bigg|{\partial\over\partial t}\,{1\over2}\int{\bf B}^2\,dV +
  R_B^{-1}\int{\bf J}^2\,dV\bigg|\;\bigg/\; R_B^{-1}\int{\bf J}^2\,dV},
\end{equation}
then $q$ never exceeded 0.01, with typical values being much smaller still.
For example, if we consider not the maximum values, but instead the rms
values over a given run, then $q_{\rm rms}$ never exceeded 0.001.

\section{Results}

\subsection{Initial Condition ${\bf B}_{p1}$}

Following Shalybkov \& Urpin (1997), we start with the simplest possible
initial condition, namely just the lowest poloidal decay mode ${\bf B}_{p1}$.
Figure 2 shows how the first three harmonics $b_1$, $b_3$ and $b_5$ of the
external field then evolve in time, where these $b_l$ are defined by
\begin{equation}
B_r(r_o,\theta,t)=\sum_l\, b_l(t)\, P_l(\cos\theta).
\end{equation}
That is, $b_l(0)$ is nothing more than the coefficient of ${\bf B}_{pl}$ in
our initial condition.  And indeed, we note how $b_1$ starts out at 1, and
then slowly decays.  It does not decay monotonically, but never deviates
very much from the $\exp(-49 R_B^{-1}\,t)$ rate that Ohmic decay alone would
have yielded.  For these runs at least, the inclusion of Hall drift
has not significantly changed the decay rate.

\begin{figure}
\psfig{file=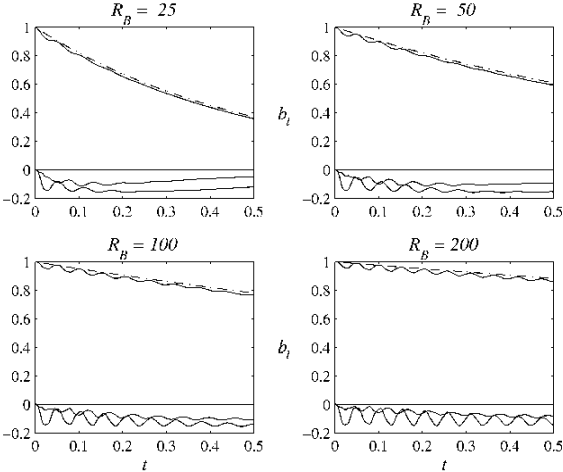,width=8.0cm}
\caption{The harmonics $b_1$, $b_3$ and $b_5$ as functions of time, for the
four indicated values of $R_B$.  The solid line starting at 1 is $b_1$, with
the dashed line being the $\exp(-49 R_B^{-1}\,t)$ decay rate of Ohmic decay
alone.  The next largest solid line is $b_3$, and the smallest $b_5$.}
\end{figure}

That is not to say that Hall drift has no influence on the field though; we
note how $b_3$ oscillates, on a timescale of approximately 0.05, and reaching
amplitudes as large as 0.15, with both the period and the amplitude largely
independent of $R_B$.  Converting back to dimensional time therefore, we
could expect periods on the order of
\begin{equation}
T\sim 0.05 \times  {10^8\over B_{12}}\;{\rm years},
\end{equation}
or $O(10^5)$ years for the very strongest fields.  These so-called helicoidal
oscillations are in excellent agreement with those previously obtained by
Shalybkov \& Urpin, who went on to derive an associated dispersion relation,
verifying that one should obtain waves that oscillate on the $O(1)$ Hall
timescale and decay on the $O(R_B)$ Ohmic timescale, exactly as we see here.

Based on these results therefore, one would think that the solution ought to
exist for arbitrarily large $R_B$, with the only effect of ever larger values
being to postpone to ever larger times the decay of both the main field $b_1$
and these oscillations in the induced field $b_3$.  Well, perhaps such a
solution does exist for arbitrarily large $R_B$, but we certainly could not
obtain it numerically.  Every attempt to increase $R_B$ much beyond 200 failed,
even at truncations as high as $120\times30$ and timesteps as small as
$10^{-8}$.

\begin{figure}
\psfig{file=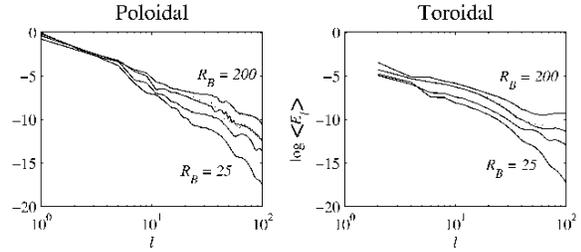,width=8.0cm}
\caption{Log$<\!\!E_l\!\!>$ versus $\log l$, where the $<>$ brackets indicate
an average over $t$ between 0.4 and 0.5 -- long enough to average over these
helicoidal oscillations, but short enough to be largely unaffected by the
overall decay.  The (barely visible) dotted lines show $R_B=100$ at
truncation $70\times20$.}
\end{figure}

We can perhaps begin to understand why by considering the power spectra shown
in figure 3.  Here
\begin{equation}
E_l={1\over2}\int{\bf B}_l^2\,dV
\end{equation}
is the energy contained in the $l$-th spherical harmonic mode, either poloidal
or toroidal, and as in (24) the integration over $r$ includes the energy in the
exterior vacuum field.  (And of course the cross-terms $\int{\bf B}_{l_1}\cdot
{\bf B}_{l_2}\,dV$ vanish by the orthogonality of the spherical harmonics.  The
total energy is thus indeed just the sum of these individual $E_l$.)

Turning
to the poloidal spectra first, we note that the $R_B=200$ curve follows
an $l^{-5}$ scaling over the entire range of $l$, whereas the lower $R_B$
curves start out much the same, but then drop off somewhat more rapidly,
exactly as one might expect.  We note though that there is no sign of a
definite dissipation scale, either at the $l=O(R_B)$ appropriate to an $l^{-2}$
spectrum, or the $l=O(R_B^{2/5})$ appropriate to an $l^{-5}$ spectrum. As
discussed in section 2.2, this suggests that the coupling is not purely local
in wavenumber space.  We note also that this particular exponent $-5$ is rather
different from the $-2$ predicted by Goldreich \& Reisenegger (1992), but of
course one should hardly expect the two to be the same, given that their result
applies to 3D turbulence, whereas our calculations here are 2D laminar.

Turning to the toroidal spectra next, for small $l$ they too are of the form
$l^p$, but now the exponent is around $-3.5$ rather than $-5$ or $-2$.  The
entire curves also shift upward slightly with increasing $R_B$, and show no
sign of saturating for sufficiently large values.  Probably more worrisome
though is the behaviour for large $l$, where the $R_B=200$ curve actually
rises ever so slightly between $l=60$ and 100.  However, runs done at
truncations varying between 80 and 120 all showed this same minimum at $l=60$,
suggesting that it is real, and not some numerical artifact.  Furthermore, the
$R_B=100$ and 50 curves also show slight rises but still fall again thereafter,
so perhaps the $R_B=200$ curve would too, if only we could include enough
modes.  It is nevertheless not quite clear what to make of this $R_B=200$
curve, and whether it really is fully resolved at the truncations we can
afford.  Based on these spectra though, we can certainly understand why
attempting to increase $R_B$ further still was not successful.

\begin{figure}
\psfig{file=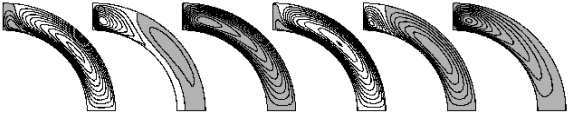,width=8.0cm}
\caption{Contour plots of the $R_B=200$ toroidal field at times 0.4, 0.42,
$\ldots$, 0.5, with a contour interval of 0.01, and negative regions
grey-shaded.  The poloidal field is not shown, as it is essentially unchanged
from ${\bf B}_{p1}$ (which is only to be expected if $b_3$ never exceeds
0.15).}
\end{figure}

Briefly returning also to the results of Shalybkov \& Urpin, we have already
noted that they too obtained helicoidal oscillations much like those in figure
2.  Unfortunately, they did not plot power spectra at all, but simply stated
that only the $l\le5$ modes ``give an appreciable contribution,'' without
further comment on what that means quantitatively.  With 40 latitudinal finite
difference points they were actually resolving considerably more than just the
$l\le5$ modes though -- although of course considerably less than the 100 modes
we have resolved here.  It is nevertheless surprising that they obtained such
good results with such a low resolution.  (In contrast, the fact that they
worked in a full sphere rather than a thin shell makes very little difference;
we also did a few runs with $r_i/r_o=0.5$ and 0.25, and obtained spectra quite
similar to those in figure 3.)

Finally, we would like to know what the solutions actually look like, and in
particular see whether we can identify the features corresponding to these
ever flatter spectra.  Figure 4 shows the field for $R_B=200$ and $t$ between
0.4 and 0.5, that is, covering the last two of these helicoidal oscillations
in figure 2 (and also precisely the time over which the spectra in figure 3 
were averaged).  We see that these oscillations involve reversals
in the sign of $B$, originating at the equator and propagating to the poles.
What we do not see, however, are any small scale features corresponding to
this part of the spectrum between 60 and 100.  In retrospect this is
probably not surprising though, since this plateau is after all 6 orders of
magnitude down from the large scale features, and therefore shouldn't be
expected to be visible on a simple contour plot such as this.  In some of our
solutions below though we will see small scale features as well, at which
point we will better understand why they break down for sufficiently large
$R_B$.

\subsection{Initial Conditions ${\bf B}_{p1} + a\,{\bf B}_{t2}$}

The maximum toroidal field in figure 4 is only 0.1, and even in the earlier
stages of evolution it never exceeds 0.25.  It is therefore probably not
surprising that $b_3$ never becomes comparable with $b_1$, since according to
(12) the toroidal field is a crucial ingredient in inducing higher harmonics
in the poloidal field.  If we did have a larger toroidal field though, it
seems likely we would also obtain a larger $b_3$, perhaps even comparable
with $b_1$.  To test this hypothesis, we add some constant $a$ times
${\bf B}_{t2}$ to our previous initial condition ${\bf B}_{p1}$.  This choice
is also consistent with the well-known fact that most dynamo models generate
toroidal fields at least as strong, if not stronger, than the poloidal field.
If we're assuming that our initial condition was generated by some previously
acting dynamo, it therefore seems reasonable to take an initial toroidal
field at least as strong as our initial poloidal field, that is, $|a|\ge1$
(and from section 2.7 we remember that here we will indeed have to consider
$\pm a$ separately).

Figure 5 shows the results for $|a|=1$ and $R_B=25$ and 50.  We obtain the
same helicoidal oscillations as before, on much the same $\sim0.05$ timescale
as before.  Now, however, the maximum amplitude of $b_3$ is indeed greater,
around 0.2 for $a=1$ and 0.3 for $a=-1$.  Except for such minor quantitative
details, the only other effect of reversing the sign of $a$ though is to
reverse the initial deflection of $b_3$ (and of $b_1$ as well).  And again as
before, the only effect of increasing $R_B$ that is evident here is to delay
the inevitable decay of both the main field and the induced oscillations.

Figure 6 shows the results for $|a|=2$, $R_B=50$, and for $|a|=3$, $R_B=25$.
And not surprisingly, the maximum amplitude of $b_3$ is greater still,
almost 0.5 for $|a|=2$, 0.62 for $a=-3$, and 0.76 for
$a=3$.  Based on these results, therefore, it seems that one should be able
to make the maximum amplitude of $b_3$ arbitrarily large, simply by taking $a$
sufficiently large.  As before though, all attempts to increase $a$ (and/or
$R_B$) much beyond the values in figure 6 failed, no matter how small a
timestep was tried.  And not surprisingly, the reason for this failure is
also much as before.  Figure 7 shows power spectra for $R_B=25$ and $a=1$, 2,
3, and we note that increasing $a$ also causes the spectra to become
flatter and flatter, until by $a=3$ the poloidal spectrum falls off by only
7 orders of magnitude, and the toroidal by 6.

\begin{figure}
\psfig{file=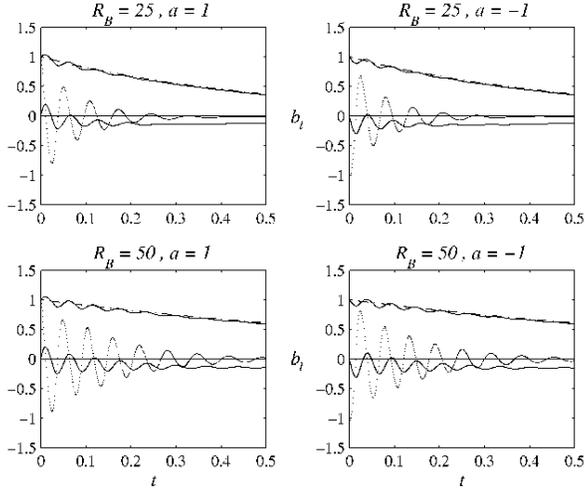,width=8.0cm}
\caption{As in figure 2, the solid line starting at 1 is $b_1$ as a function
of $t$, with the dashed line again being the $\exp(-49 R_B^{-1}\,t)$ decay
rate of Ohmic decay alone.  The solid line starting at 0 is again $b_3$.  The
dotted line starting at $\pm1$ is the toroidal field at $r=0.875$, $\theta=
\pi/4$, where ${\bf B}_{t2}$ takes its maximum value.}
\end{figure}

\begin{figure}
\psfig{file=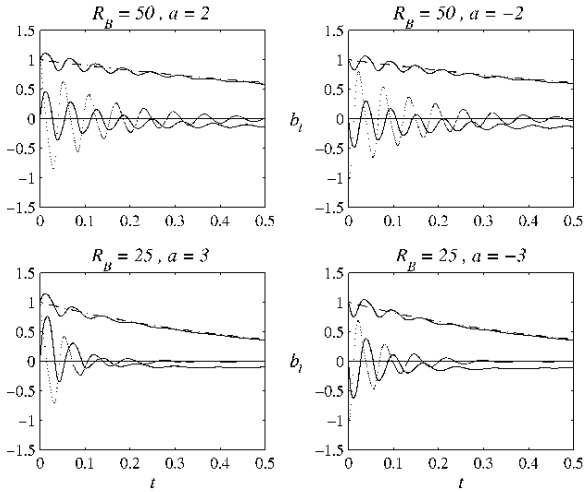,width=8.0cm}
\caption{As in figure 5, except that now the dotted line is $B(0.875,\pi/4)
/|a|$.  We see therefore that increasing $a$ simply scales up the toroidal
field, but otherwise has virtually no effect on it, only on $b_3$.}
\end{figure}

The toroidal spectrum therefore drops off by 6 orders of magnitude for both
the $R_B=200$ run above, as well as for this $a=3$ run here.  Here though
the dropoff is much more uniform between $l=2$ and 100, whereas above we
remember it all occurred between 2 and 60.  That necessarily means then
that this run has more power in the intermediate range.  If we are therefore
trying to identify the features corresponding to these ever flatter spectra,
this is the run to consider.  Figure 8 shows the field up to $t=0.06$, so the
first of these oscillations this time.  We note how they
again involve reversals in the sign of $B$.  This time, however, we can also
see some small scale structures emerging; at $t=0.025$ and again between
0.05 and 0.06 some of the contour lines near the equator crowd very close
together, indicating the formation of very intense currents.
(Some of the contour lines near the inner boundary also crowd very close
together, suggesting that our $B=0$ boundary condition -- which we remember
depended on $\partial B/\partial r$ being small -- should be viewed as a
mathematically convenient toy model rather than a physically accurate
approximation.)

\begin{figure}
\psfig{file=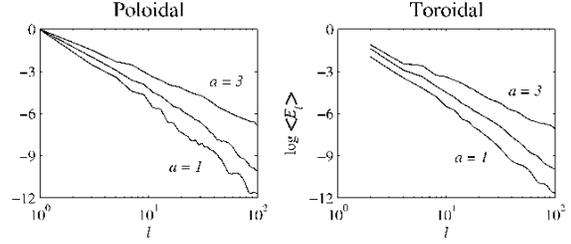,width=8.0cm}
\caption{Power spectra of the solutions for $R_B=25$ and $a=1$, 2 and 3, this
time averaged between $t=0$ and 0.1 to avoid the subsequent very strong decay.
As one might expect, the spectra for $R_B=50$, $a=1$ and 2, are flatter than
the corresponding ones for $R_B=25$, but still not as flat as the $a=3$ ones
shown here.  We therefore show only the $R_B=25$ spectra, to focus attention
on the variation with $a$.  We also note how all of these spectra are roughly
of the form $l^{p}$, with $p$ varying between $-3.5$ and $-6$ for both poloidal
and toroidal.}
\end{figure}

\begin{figure}
\psfig{file=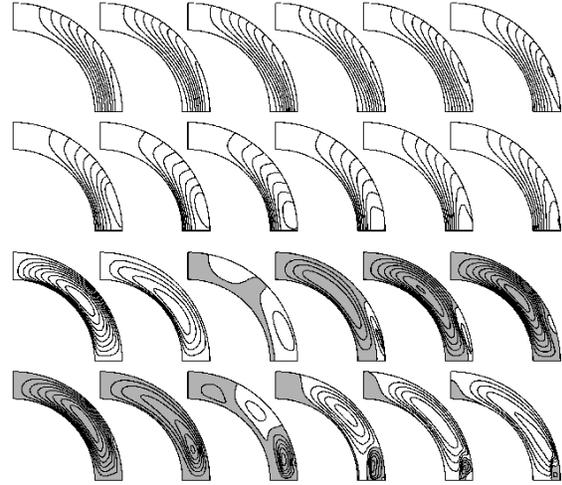,width=8.0cm}
\caption{The structure of the $R_B=25$, $a=3$ field, at times 0.005, 0.01,
$\ldots$, 0.06.  The top two rows show the poloidal field, where we note
that now we do see significant departures from ${\bf B}_{p1}$, namely the
emergence of field lines closing entirely in one hemisphere.  The bottom
two rows show the toroidal field, with a contour interval of 0.25, and
negative regions again grey-shaded.}
\end{figure}

So here we have the small scale structures corresponding to the ever flatter
spectra, and therefore also the reason we cannot increase $R_B$ and/or $a$
indefinitely.  As we increase these parameters, these structures get finer
and finer, until we can no longer resolve them, and the code inevitably 
crashes.  The only remaining questions then are precisely how the thickness of
these structures scales with $R_B$, and whether the governing equations always
remain parabolic, or whether some critical $R_B$ is eventually reached beyond
which they switch to hyperbolic at various points in space and time.  What
would happen after that is completely unknown, of course, and unfortunately
not answerable with this code.

\subsection{Initial Condition ${\bf B}_{p2}$}

Figure 9 shows the results starting with ${\bf B}_{p2}$ as the initial
condition.  Comparing with figure 2, we see that one obtains far larger
oscillations in the higher harmonics, with even $b_8$ still attaining a
quite substantial amplitude.  The period is also longer, 0.2 instead of 0.05.
As before though, both the period and the amplitude are only weakly dependent
on $R_B$, and even that probably only because here we cannot achieve
sufficiently high values to have much more than one cycle before everything
decays.  And we note, incidentally, that now the main field $b_2$ does decay
slightly faster than Ohmic decay alone would have dictated, but still not
enough to be significant.  Finally, figure 10 shows the field through the
first of these oscillations, and again we see the emergence of very small
scale structures at certain times.

\begin{figure}
\psfig{file=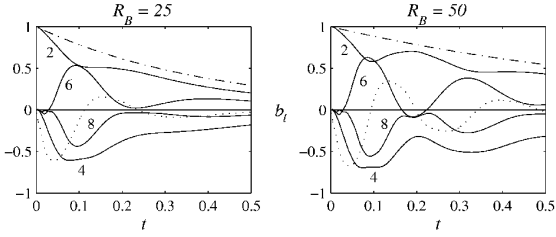,width=8.0cm}
\caption{The harmonics $b_2$ through $b_8$ as functions of time.  The dashed
line is now the $\exp(-61 R_B^{-1}\,t)$ decay rate of $b_2$ if Ohmic decay
alone were acting.  Finally, the dotted line is again $B(0.875,\pi/4)$.}
\end{figure}

\begin{figure}
\psfig{file=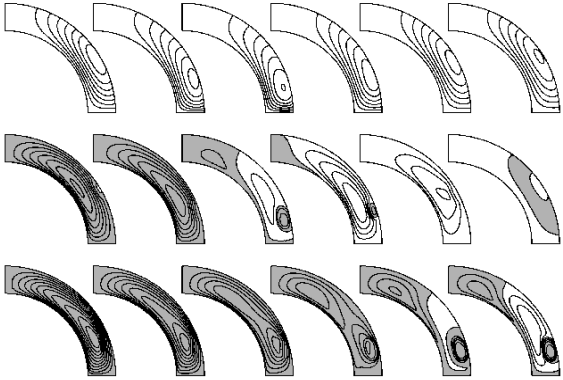,width=8.0cm}
\caption{The structure of the $R_B=50$ field.  The top two rows show the
poloidal and toroidal fields at times 0.033, 0.067, $\ldots$, 0.2, with
contour intervals of 0.025 and 0.1, respectively.  The bottom row shows the
toroidal field at times 0.075, 0.08, $\ldots$, 0.1, with a contour interval
of 0.05, in order to see the emergence of this very intense current
loop in more detail.}
\end{figure}

\subsection{Initial Condition $2/3\,{\bf B}_{p1}+1/3\,{\bf B}_{p2}$}

All of the solutions presented so far have belonged to one or the other of the
two equatorial symmetry classes discussed in section 2.7.  To get at least
some idea of what might happen when these two families are allowed to
interact, we (rather arbitrarily) take the initial condition $2/3\,{\bf B}_{p1}
+1/3\,{\bf B}_{p2}$ (so $B_r(r_o,0)=1$ is still the maximum surface field).
Figures 11 and 12 show these results.  A number of interesting features to note
are how $b_1$ is still almost completely unaffected by the inclusion of Hall
drift, whereas $b_2$ is so strongly affected that it now oscillates in sign,
rather than decaying monotonically as in figure 9.  The higher harmonics $b_3$
and $b_4$ are also excited, with maximum amplitude around 0.2 for both.

One could now obviously do endless more runs, for example to discover how
large the initial $b_1$ must be to cause $b_2$ to oscillate, or whether a
sufficiently large initial $b_2$ could cause $b_1$ to oscillate instead.
However, given that there is no observational or theoretical reason to prefer
any of these linear combinations over any other, that seems rather pointless.
It is already evident in figure 11 that Hall drift affects mixed parity
solutions in much the same way as pure parity solutions, and that is probably 
all we can expect to learn from these runs.

\begin{figure}
\psfig{file=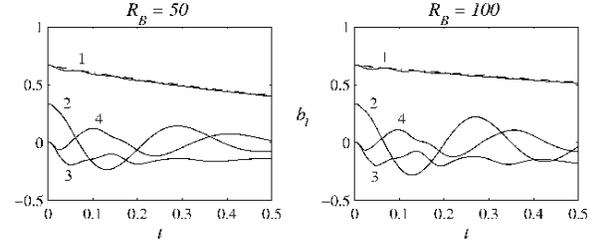,width=8.0cm}
\caption{The harmonics $b_1$ through $b_4$ as functions of time.}
\end{figure}

\begin{figure}
\psfig{file=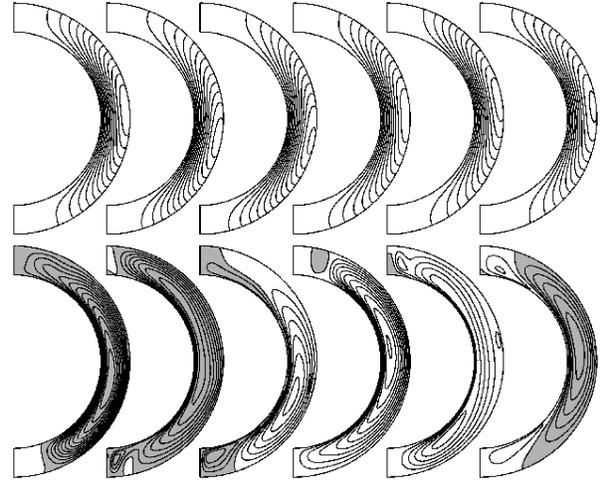,width=8.0cm}
\caption{The structure of the $R_B=100$ field, at times 0.05, 0.1, $\ldots$,
0.3, with contour intervals of 0.025 and 0.05, respectively.  One interesting
feature to note here is how the toroidal field is dominated by the
${\bf B}_{t1}$ mode, the one that is absent in either of the pure parity
families.}
\end{figure}

\section{Conclusion}

The results presented here suggest that Hall drift could indeed have a
significant influence on the evolution of a neutron star's magnetic field.
Particularly if the internal toroidal field is as strong or stronger than the
poloidal field, Hall drift can excite some of the higher harmonics to
amplitudes comparable to the original mode.
However, as substantial as some of these higher harmonics are, this still does
not appear to be enough to cause the original mode to decay significantly
faster than it otherwise would have.  This conclusion must be qualified though
by our inability to increase $R_B$ indefinitely.  Indeed, the very feature
that caused the code to fail beyond certain limits, namely the fact that the
spectra got flatter and flatter, also indicates that this transfer of energy
to the higher harmonics gets more and more efficient as $R_B$ is increased.
It is conceivable, therefore, that the solutions for, say, $R_B=1000$ would
show a very rapid decay of the original mode.  Also, the cascade may well be
very different in 3D than in 2D, just like ordinary fluid turbulence is very
different.  Extending our model here from 2D to 3D is possible in principle,
but will obviously require considerable computational resources.

Finally, even if it should turn out that Hall drift alone, in either 2D or 3D,
simply does not generate a sufficiently strong cascade at any value of $R_B$,
the combination of Hall drift and stratification may still do so.  We've
already noted in the introduction that the electron number density $n$ in
equation (2) is in fact not constant, but rather varies by several orders of
magnitude over the depth of the crust.  Vainshtein et al.\ (2000) show that if
one includes this effect, one can obtain a very rapid decay of a toroidal field
at least.  In their highly idealized analytical model it was not possible to
include poloidal fields though (we recall from section 2.4 that one can
indeed consistently consider only toroidal fields).  In contrast, our numerical
model here already includes poloidal fields, and including radial variations in
$n$ is possible too.  Calculations are therefore currently under way to see if
this Vainshtein et al.\ result applies to poloidal fields as well.

\section*{Acknowledgments}

RH's stay in Germany was made possible by a Research Fellowship of the
Alexander von Humboldt Foundation.

\label{lastpage}

\end{document}